\documentclass[11pt,preprint]{aastex}
\usepackage{natbib}
\usepackage{epstopdf}

\slugcomment{Submitted to PASP}

\shorttitle{Planar Detectors in the Few-Mode Limit}
\shortauthors{Chuss et al.}

\begin{document}


\title{Diffraction Considerations for Planar Detectors in the Few-Mode Limit}

\author{David T. Chuss, Edward J. Wollack, S. Harvey Moseley}
\affil{NASA Goddard Space Flight Center, Observational Cosmology Laboratory, Code 665,
Greenbelt, MD, 20771, email contact: David.T.Chuss@nasa.gov}

\and

\author{Stafford Withington, George Saklatvala}
\affil{University of Cambridge, Cavendish Laboratory, JJ Thomson Avenue, Cambridge CB3 OHE}

\begin{abstract}
Filled arrays of bolometers are currently being employed for use in astronomy from the far-infrared
through millimeter parts of the electromagnetic spectrum. Because of the large range of wavelengths for
which such detectors are applicable, the number of modes supported by a pixel will vary according to the specific
application of a given available technology. We study the dependence of image fidelity and induced polarization on the size of the pixel by employing a formalism in which diffraction due to the pixel boundary is treated
by propagating the second-order statistical correlations of the radiation field through a model optical system. 
We construct polarized beam pattern images of square pixels for various ratios of $p/\lambda$ where $p$ is the pixel size and $\lambda$ is the wavelength of the radiation under consideration. 
For the limit in which few modes are supported by the pixel ($p/\lambda<1$), 
we find that the diffraction due to the pixel edges is non-negligible and hence must be considered along with the telescope
diffraction pattern in modeling the ultimate spatial resolution of an imaging system.
For the case in which the pixel is 
over-moded ($p/\lambda>1$), the geometric limit is approached as expected. 
This technique gives a quantitative approach to optimize the imaging properties of arrays of planar detectors in
the few-mode limit.
\end{abstract}

\keywords{techniques: polarimetric --- techniques: photometric --- methods: statistical}

\section{Introduction}

Filled arrays of planar detectors have a long history of astronomical use in the optical (CCDs) and infrared (InSb, HgCdTel) 
regions of the electromagnetic spectrum. In recent years, planar bolometers have enabled the extension of this 
technique into the millimeter range where coherent detection techniques have traditionally been used.  
Examples of planar millimeter-wave elements can include both patches of absorbing material, which can be multimode if made large enough, or single-mode antenna-coupled structures. In this paper, we are concerned with detectors that operate through the absorption of radiation in a lossy thin-film or bulk material.

The constituent detectors of planar arrays are normally considered in the geometric limit. In other words, 
the coherence length associated with the loss mechanism that is responsible for power absorption
is small compared with the detector's physical size. In this limit, the only diffractive effects considered 
are those induced by the optical system that controls the illumination of the detectors, and the individual 
pixels of the array merely extract power from the incoming field according to their geometrical size.  
In such a model, the sampling of a telescope-induced Airy disk is assumed to be perfect, and the Airy disk 
sets the angular resolution limit  of the optical system as long as there
are a sufficient number of pixels within the disk. 

In order for the detector to behave in a 
geometrical manner, the ratio of the pixel size ($p$) to the wavelength ($\lambda$) must be significantly greater than 
unity. In many cases, however, the individual pixels are nearly a wavelength in size, and so their behavior cannot be described in terms of geometrical shape alone, as the pixel boundary conditions in this limit become important. For example, the currents flowing around the edges of the individual pixels will lead to beam patterns that exhibit diffractive
phenomena, including potentially complicated states of polarization. Another way of describing the change 
from geometrical to diffractive behavior is that the number of electromagnetic modes that can be supported
by a detector decreases as the size of the detector decreases. 

In millimeter-wave astronomy, practical limitations such as the size of the optical system and those 
associated with physical properties relating to the speed of bolometric detectors (thermal conductivity and 
heat capacity), motivate the construction of fast optical systems that have lower $p/\lambda$ 
than their high-frequency counterparts. In this case, the diffractive effects of the pixels and their associated 
architecture are no longer negligible, and the consequences for the imaging capability of the system must be understood. 
Such image degradation is analogous to that of infrared photodiode arrays 
\citep{Holloway86}; however, in this case, since $1<p/\lambda<5$, the image degradation is not due to diffractive 
effects, but rather to diffusive spreading of charge carriers in a common semiconductor absorber. In both 
cases, these focal plane effects must be considered when designing the front-end optics. A partial list of current 
and planned instruments that employ planar bolometers is given in Table~\ref{tab:instruments}. These include 
HAWC/SOFIA \citep{Harper04}, SHARC II/CSO \citep{Dowell03}, SCUBA 2/JCMT \citep{Holland06}, PAR/GBT \citep{Dicker06}, 
GISMO/IRAM \citep{Staguhn06}, and ACT \citep{Fowler04}.

In this work, we explore the diffractive effects of pixels for which $0.25\le p/\lambda \le 4.00$. We adapt a 
method in which the statistical correlations of an incoming radiation field, as distinct from the field itself, 
are propagated through a model optical system \citep{Withington06, Withington03}. Care has been taken in the 
simulations to ensure that the diffractive effects seen on the sky are dominated by the pixels and not by 
the optical system that illuminates the array. This has been done with the specific purpose of isolating the 
effects of using small pixels from the effects of the telescope optics. We find that when $p/\lambda >1$, the geometric 
limit is recovered, but that when $p/\lambda <1$, diffractive effects are non-negligible, and must be taken into 
account when designing instruments that employ such pixels. 

In Section~2, we briefly review the method of \cite{Withington03} as it applies to our calculations. In Section~3, 
we describe the results of the calculated polarized and unpolarized beam patterns for various values of $p/\lambda$. 
In Section~4, we suggest a method for using the beam patterns to calculate the cross-coupling between pairs 
of pixels when viewing an extended, incoherent source. In Section~5, we briefly consider the beam shape of a 
polarized pixel. Finally, conclusions and future work are discussed in Section~6.  

\section{The Method}

Sections 2.1 and 2.2 are provided as a brief summary of \cite{Withington03}. For a more thorough discussion
of the method they have developed, the reader is invited to consult their work.  In Section 2.3 we connect
the correlation dyadic and the Stokes parameters. We later employ this connection in our analysis and presentation
of the diffractive effects of the pixels. Finally, Section 2.4 concludes with a discussion of the number of modes
required in order to ensure that the diffraction in the model system is dominated by the pixels and not due to other
optical elements in the system.

\subsection{Basic Formalism}

The correlations between the components of an electric field at two locations ($\overline{r}_1$ and $\overline{r}_2$) 
are conveniently expressed in terms of a space-domain correlation dyadic:
\begin{equation}
\overline{\overline{E}}(\overline{r}_1,\overline{r}_2)=\left\langle\overline{E}(\overline{r}_2)\overline{E}^*(\overline{r}_1)\right\rangle.
\end{equation}
The angle brackets denote an ensemble average. 

The space-domain correlation dyadic encodes the second-order statistics of the radiation field. For purposes
of this work, we take the $z$-axis to be the axis of the optics. The second-order statistics can be propagated
from one plane ($z=constant$) to another in the optical system by implementing the appropriate boundary conditions.

\begin{figure}[h]
\begin{center}
\includegraphics[width=3.0in]{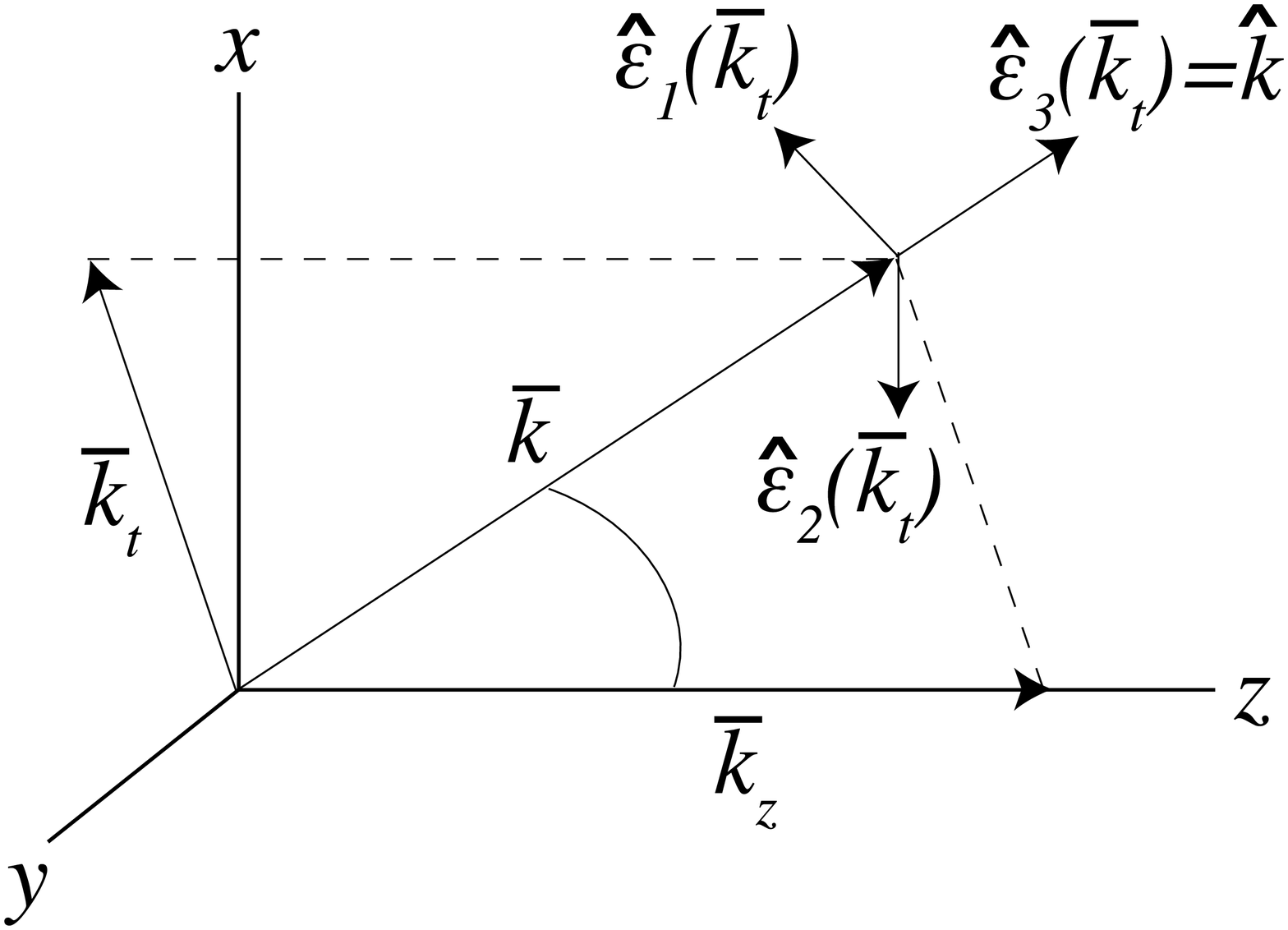}
\caption{The coordinate system for the formalism is shown. The $(x,y,z)$ coordinate system is that of the optical system
with $z$ being the optic axis. The $(\hat{\varepsilon}_1({\overline{k}_t)},\hat{\varepsilon}_1({\overline{k}_t)},
\hat{\varepsilon}_1({\overline{k}_t)})$ coordinate system is that corresponding to the plane wave mode characterized by
$\overline{k}_t$.  The plane of constant phase is spanned by $\hat{\varepsilon}_1({\overline{k}_t)}$ and $\hat{\varepsilon}_2({\overline{k}_t)}$, and the direction of propagation is given by $\hat{\varepsilon}_1({\overline{k}_t)}=\hat{k}$. }
\label{fig:coordsys}
\end{center} 
\end{figure}

Following \cite{Withington03},
the electric field can be decomposed into a complete, orthogonal set of vector plane-wave fields,
\begin{equation}
\overline{\psi}_i(\overline{k}_t,\overline{r}_t)=\frac{1}{2\pi}\hat{\varepsilon}_i(\overline{k}_t)e^{j\overline{k}_t\cdot\overline{r}_t}
e^{jk_zz}e^{-j\omega t},
\end{equation}
where, $\overline{k}=k\hat{k}=k_x\hat{x}+k_y\hat{y}+k_z\hat{z}=\overline{k}_t+k_z \hat{z}$, and $\hat{\varepsilon}_i(\overline{k}_t)$ 
is the set of unit polarization vectors in which $\hat{\varepsilon}_3(\overline{k}_t) = \hat{k}$, the direction normal to 
the planes of constant phase.  Figure~\ref{fig:coordsys} illustrates the relationships between the relevant vectors.

Given  this basis, and denoting the Fourier amplitude for the mode characterized by transverse wave vector $\overline{k}_t$ for the $i$th polarization as $a_i(\overline{k}_t)$, the electric field can be expressed in as
\begin{equation}
\overline{E}(\overline{r})=\frac{1}{2\pi}\int \sum_{i=1}^3 a_i(\overline{k}_t)\hat{\varepsilon}_i(\overline{k}_t)e^{j\overline{k}_t\cdot\overline{r}_t}e^{jk_zz}e^{-j\omega t}d^2\overline{k}_t.
\end{equation}
It is convenient, and equivalent, to express the correlations of the electric field in terms of the $k$-domain 
correlation dyadic,
\begin{equation}
\overline{\overline{A}}(\overline{k}^\prime_t,\overline{k}_t)=\sum_{i=1}^3\sum_{j=1}^3 \left\langle 
a_i(\overline{k}_t)a_j^*(\overline{k}^\prime_t)\right\rangle \hat{\varepsilon}_i(\overline{k}_t)\hat{\varepsilon}_j(\overline{k}_t^\prime).
\end{equation}
Because
we are modeling detectors that have a planar geometry, we are concerned with correlations between points in
the same plane. That is, $z_1=z_2=0$. Defining $\overline{r}_{t1}=x_1\hat{x}+y_1\hat{y}$ and $\overline{r}_{t2}=x_2\hat{x}+y_2\hat{y}$, the space- and $k$-domain correlation dyadics are related by two 2-dimensional Fourier transforms.
\begin{eqnarray}
\overline{\overline{A}}(\overline{k}^\prime_t,\overline{k}_t)=\frac{1}{(2\pi)^2}\int\int 
\overline{\overline{E}}(\overline{r}_1,\overline{r}_2) e^{-j\overline{k}_t\cdot\overline{r}_{t2}}e^{j\overline{k}^\prime_t\cdot
\overline{r}_{t1}}e^{-jk_zz_2}e^{jk^\prime_z z_1}d^2\overline{r}_{t1}d^2\overline{r}_{t2}\label{eq:EtoA}\\
\overline{\overline{E}}(\overline{r}_1,\overline{r}_2)=\frac{1}{(2\pi)^2}\int\int 
\overline{\overline{A}}(\overline{k}^\prime_t,\overline{k}_t) e^{j\overline{k}_t\cdot\overline{r}_{t2}}e^{-j\overline{k}^\prime_t
\cdot\overline{r}_{t1}}e^{jk_zz_2}e^{-jk^\prime_z z_1}d^2\overline{k}_{t}d^2\overline{k}^\prime_{t}\label{eq:AtoE}
\end{eqnarray}

\subsection{Model Optical System}

\begin{figure}[h]
\begin{center}
\includegraphics[width=5.0in]{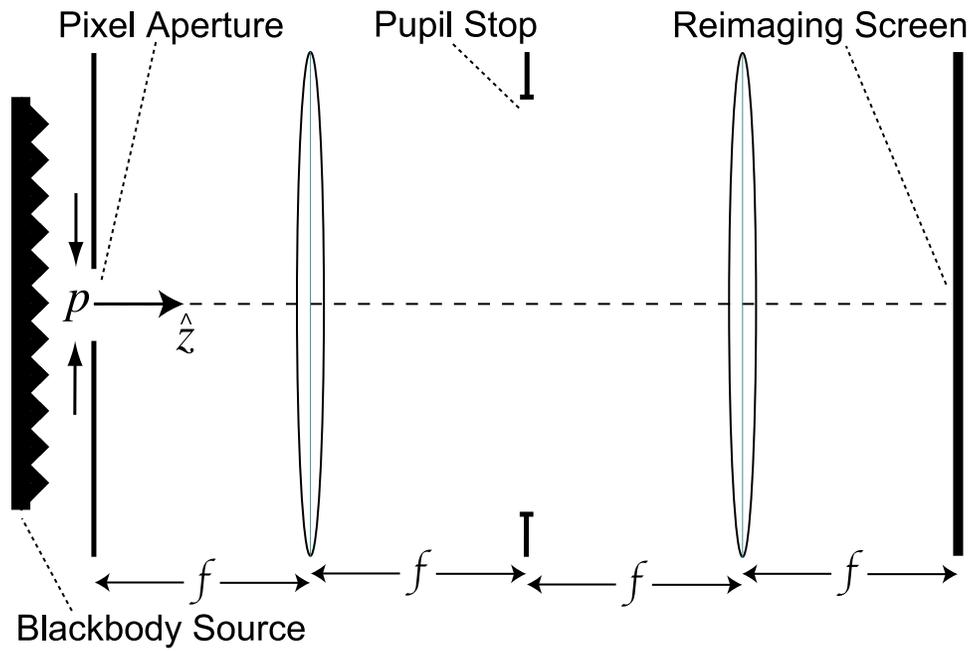}
\caption{The model optical system that is used for the analysis is shown. The blackbody source produces radiation that gets scattered by the 
pixel aperture. The pupil stop is made large enough to pass all of the modes produced by the blackbody, and the radiation pattern is imaged on the reimaging
screen.}
\label{fig:sysdia}
\end{center} 
\end{figure}

The above formalism is useful for analyzing diffraction in a generic optical system.  
A diagram of the system used to study the behavior of planar bolometer arrays is shown in 
Figure~\ref{fig:sysdia}, which is a modified version of Figure~1 of Withington et al. (2003). In this 
model, a pixel is represented by a square aperture that is illuminated from the back by a blackbody
source of infinite spatial extent. The aperture scatters the radiation from the source, thereby 
introducing additional correlations into the reception pattern. The aperture plane is then imaged 
at the reimaging screen. The reimaging screen in this setup represents the projection of the pixel on the
sky.  

A key assumption is that it is appropriate to model a planar absorber as a 
hole in front of a blackbody emitter. It is our belief, that for the purposes of understanding general behavior,
this assumption is quite valid, since to the extent that the absorber is black, the behavior of the system is insensitive
to the physical realization of the detector. Various small changes can be made to take into account the 
details of a specific absorber, but a generic model
that makes no assumptions about the precise construction of the pixels, but which brings out key
aspects of behavior, is highly beneficial. Future simulations can be much more sophisticated.

Because the size of the whole imaging array is finite in practice, we work with a discrete basis set, 
\begin{equation}
\overline{\psi}_i(m,n;x,y)=\frac{1}{L}\hat{\varepsilon}_i(m,n)e^{j(k_x(m)x+k_y(n)y)}e^{jk_zz}e^{-j\omega t}.
\end{equation}
Here, $L$ is the size of the blackbody emitter (assumed to be square), and $m$ and $n$ are the discrete
mode numbers in the $x$ and $y$ directions, respectively. In addition, $k_x(m)=\frac{2\pi m}{L}$, 
$k_y(n)=\frac{2\pi n}{L}$, and $k^2_z(m,n)=k^2-k^2_x(m)-k^2_y(n)$. 

In this basis, the space-domain correlation dyadic for a finite blackbody source is given
by \cite{Withington03} as
\begin{equation}
E_{rs}(\overline{r}_1,\overline{r}_2)=\frac{1}{L^2}A_o(k)\sum_{m,n}\frac{(k^2\delta_{rs}-k_rk_s)}{k^2}
e^{jk_x(m)(x_2-x_1)}e^{jk_y(n)(y_2-y_1)}.
\end{equation}
Here, the indices $r$ and $s$ correspond to the components of the space-domain correlation dyadic when
projected into the $(x,y,z)$ coordinate system. 
Also, $A_o(k)$ is the amplitude of the electric field having wavenumber $k$ and $\delta_{rs}$ 
is the Kronecker delta. These spatial correlations are the combination of those inherent to 
propagating blackbody radiation \citep{Mehta64a} and those due to the sampling of the source 
\citep{Withington03}. 

At this point, the $k$-domain correlation dyadic can be calculated using Equation~\ref{eq:EtoA}.  
The effect of the pixel aperture on the $k$-domain correlation dyadic can be expressed in matrix 
notation as $\mathbf{A^\prime}=\mathbf{SAS^\dagger}$. The $k$-domain correlation dyadic to the left 
of the pixel aperture is $\mathbf{A}$ and that to the right is $\mathbf{A^\prime}$.  
The scattering matrix appears twice in the transformation because $\mathbf{A}$ consists of two electric fields, both 
of which are scattered by the aperture.

The scattering matrix elements are given by
\begin{equation}
S_{ij}(r,s;m,n)=\frac{p^2}{L^2}\hat{\varepsilon}_i(m,n)\cdot\hat{\varepsilon}_j(r,s)\,
j_o\left(\frac{\pi p}{L}(m-r)\right)
j_o\left(\frac{\pi p}{L}(n-s)\right) 
\end{equation}
Here, $j_o(x)\equiv\sin{x}/x$. 
This scattering matrix describes the mapping of an incident mode characterized by ($m$,$n$) to 
scattered mode ($r$,$s$), where $i$ and $j$ represent the scattered and incident polarizations, 
respectively. Here, $p$ is the length of a side of a square pixel. 

As discussed in \cite{Withington03}, the pupil stop can be modeled, to first order, simply by cutting 
off the number of modes allowed to propagate through the system. In fact, this simplicity is one of the 
strengths of working with the $k$-domain dyadic.  However, since we are interested in isolating the 
effect of pixel diffraction, we wish to minimize the diffractive effects of the fore optics.  Because 
of this, we ensure that the pupil passes all of the modes produced by the finite-sized detector. 
We then proceed to calculate the space-domain dyadic components using Equation~\ref{eq:AtoE} to 
produce an image of the pixel on the reimaging screen. 

\subsection{Image Fidelity and Polarization}

The space-domain correlation dyadic contains all of the information about the second-order statistics of the
radiation field at any given plane of the optical system.  In examining the correlations introduced by the optical
system at the reimaging screen (or any other plane in the optical system for which the space-domain correlation
dyadic has been calculated), it is useful to separate the correlations into those
in which $\overline{r}_1\neq\overline{r}_2$ and those in which $\overline{r}\equiv\overline{r}_1=\overline{r}_2$.
The former describes the correlation between the fields in two different locations in the plane that
will manifest itself in the form of image fidelity. The latter describes the polarization of the beam.

The polarization can be effectively treated by noticing that
the Cartesian components of the correlation dyadic can be equivalently expressed in terms of the Stokes parameters.  
\begin{eqnarray}
I= E_{xx}(\overline{r},\overline{r})+E_{yy}(\overline{r},\overline{r}) \label{eq:I}\\
Q=E_{xx}(\overline{r},\overline{r})-E_{yy}(\overline{r},\overline{r})\label{eq:Q} \\
U=\Re{E_{xy}(\overline{r},\overline{r})}\label{eq:U} \\
V=\Im{E_{xy}(\overline{r},\overline{r})}\label{eq:V}.
\end{eqnarray}
In this work, we utilize the connection between the space-domain correlation dyadic and the Stokes parameters
to characterize the diffraction induced by an optical system in terms of its polarizing
effect on initially unpolarized radiation. We are specifically interested in characterizing
both the polarization and spatial coherence introduced by the boundary of a pixel.

\subsection{Lateral Size of the System}

The goal of this work is to study the diffractive effects of the pixel edges, and so we would like our 
model optical system to highlight the correlations due to the scattering through the pixel aperture, 
while suppressing those due to other parts of the optics. Because we have disallowed mode truncation at the 
pupil stop, the only way correlations can be introduced into the system by anything other than the pixel 
is through the finite size of the blackbody source, or in other words the total size of the region needed
numerically to define the Fourier series.  We ideally want the number of modes to be high enough to 
approximate a source of infinite lateral size; however, 
the inclusion of more modes slows the calculation.  In addition, because the size of the pixel aperture 
will naturally cut off higher modes, inclusion of modes above a certain level is expected to have minimal 
impact on the result. We seek a quantitative way to verify that we have included enough modes such that 
the correlations due to the optics are small compared to those due to the pixel aperture. 

\begin{figure}[h]
\begin{center}
\includegraphics[width=5.0in]{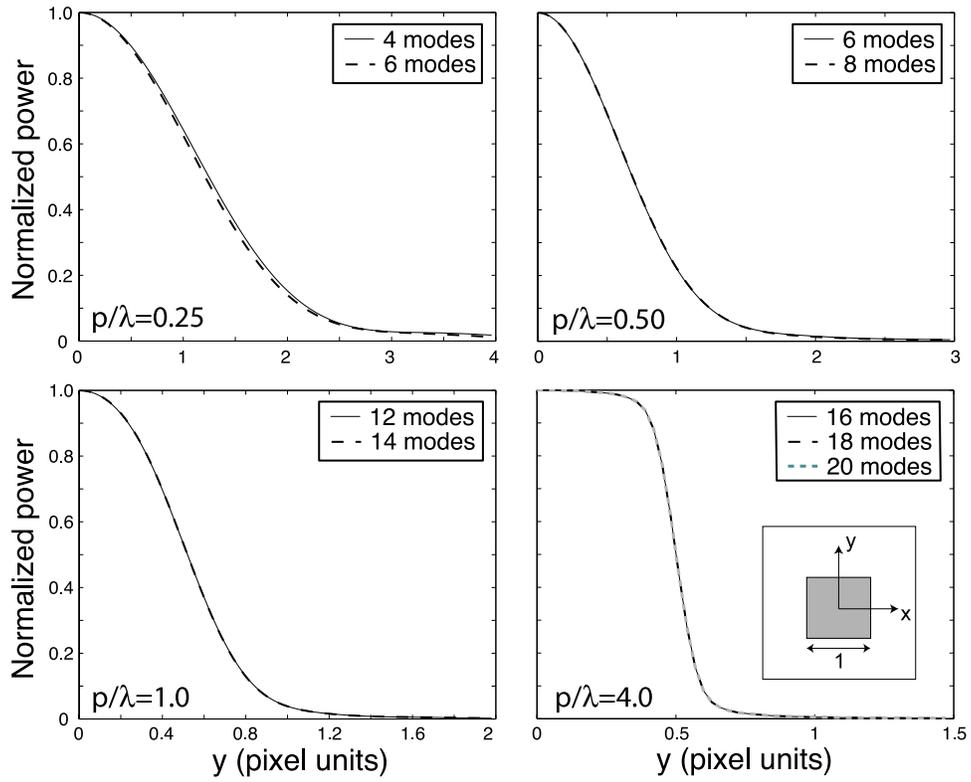}
\caption{The effect of varying the number of modes for selected cases of p/$\lambda$ is shown. The y coordinate represents
the lateral distance from the center of the pixel. The profiles plotted are those for total power (Stokes I).}
\label{fig:modeplots}
\end{center}
\end{figure}

To estimate the effect of the $k$-domain cutoff introduced by the finite size of the system, we ran 
1-dimensional calculations for Stokes I (total power) from the center to the edge of the pixel for multiple 
numbers of modes. The results of this study are shown in Figure~\ref{fig:modeplots}.  In all cases, the number of 
Fourier modes used in the calculation is significantly greater than the number of modes nominally supported
by the pixel. Based on these results, we are satisfied that the diffractive beam spreading effects that we 
calculate are due to the electromagnetic boundary conditions at the pixel edges, and not from elsewhere in 
the optical system, or from the finite size of the system used to model behavior.  The number of modes we have
included in our study of pixel diffraction is summarized in Table~\ref{tab:usedmodes}.

\section{Characterization of Pixel Diffraction}

We calculate the space-domain correlation dyadic at the reimaging screen for each of five
different values of $p/\lambda$. In order
to visualize and quantify the effect of pixel diffraction, we then calculate the spatial distribution of the 
Stokes parameters using equations~(\ref{eq:I}-\ref{eq:V}).  Both the spatial and polarization correlations
are captured in this presentation. The calculations of polarized beam patterns have the advantage that they
provide the capability for a direct comparison to measured data. They also can provide guidance when modeling predicted instrument performance. 

\begin{figure}[h]
\begin{center}
\includegraphics[width=5.0in]{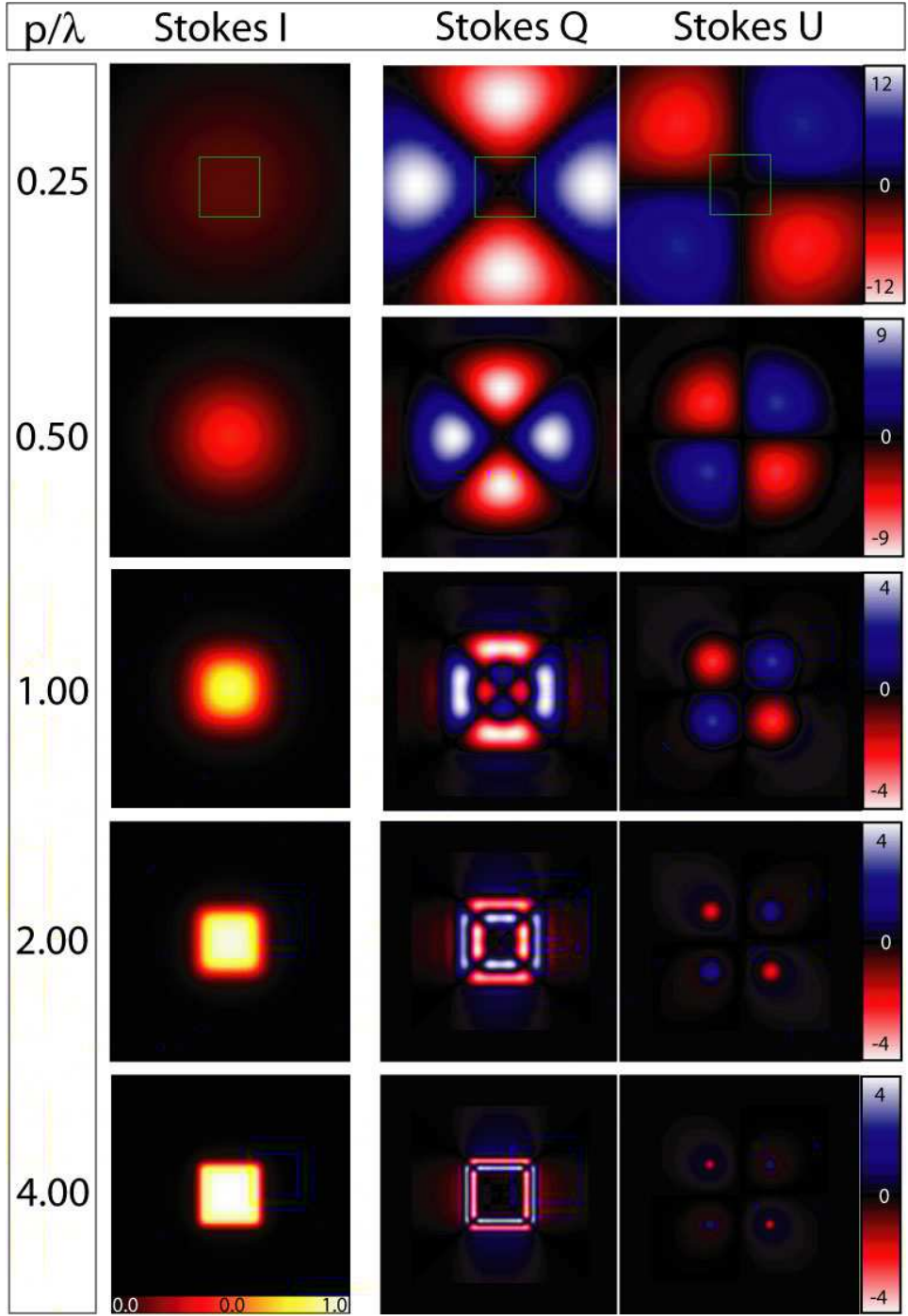}
\caption{Stokes I, Q, and U are shown for various values of $p/\lambda$.
Stokes I in each case is normalized such that the total power is constant across all five $p/\lambda$
cases such that the common color table. Stokes Q and U are normalized to the peak of each of the corresponding Stokes I flux distribution. Their corresponding color 
tables are given in per cent.
The pixel size is shown (green squares) for the
$p/\lambda=0.25$ case, and is the same size for all of the cases.}
\label{fig:Imaps}
\end{center}
\end{figure}

Figure~\ref{fig:Imaps} shows the beam patterns calculated for Stokes parameters I, Q, and U, 
for various choices of $p/\lambda$. The intensity plots are commonly normalized to the peak flux
of the $p/\lambda=4.00$ case.  Each polarization (Q and U) plot is normalized to the peak flux in the 
corresponding Stokes I plot, and colorbar units are given in per cent.  As an aside, we note that Stokes 
V is also calculable, but the symmetry of the system prevents the generation of quadrature correlations
between orthogonal polarizations in our linear basis.

In the limit of large $p/\lambda$, we expect the geometric limit to be recovered. In this case, all 
of the Stokes I flux is contained in the square that defines the physical pixel. Figure~\ref{fig:Imaps} 
shows that diffractive effects are small at $p/\lambda$=2, and the geometric limit is indeed nearly 
recovered by $p/\lambda$=4. In these cases, the diffractive effects of the pixel are more easily seen 
in the beam patterns of Stokes Q and U. For the high $p/\lambda$ cases, Stokes Q traces the regions 
close to the edges of the pixel in the following way: Immediately inside the pixel edge, the 
polarization direction is parallel to the pixel edge. Q is zero at the pixel edge and then switches 
sign such as to be perpendicular to the pixel just outside the edge. Farther out, the polarization is 
lower and parallel to the pixel edge indicating that the highly scattered radiation is polarized 
parallel to the pixel edge.  Stokes U appears at the corners 
of the pixel. Since U is defined as the in-phase correlation between vertical and linear polarization, 
the degree to which Stokes U appears depends on the spatial coherence scale for a given wavelength. 
The point-like behavior may be attributed to current flowing around the corners of the detector.
As $p/\lambda$ gets large, the area of the beam characterized by non-zero Q and/or U becomes 
increasingly small. We find that $\sim10$\% of the flux is polarized at $p/\lambda=1.0$, and it drops as 
$(p/\lambda)^{-1}$, consistent with the geometrical theory of diffraction \citep{Keller62}. 

It should be noted that this effect is more severe for U than for Q since for 
the latter, opposite signed regions are located close to one another and
are likely to cancel when convolved with a source (or the Airy disk of the telescope).
 The U beam patterns are spatially separated 
at high $p/\lambda$, and therefore are more likely to convert unpolarized anisotropies on the sky 
into polarized signals. 

For low $p/\lambda$, the coherence length is large compared to the size of the pixel.  Stokes I 
transitions from a square-like pattern to a circular beam shape. In these cases, the wavelength of
the radiation is too large to resolve the details of the pixel shape.  The effect of diffraction is
significant, as indicated by the larger spatial extent of Stokes I and the larger values and more 
extended structure of Stokes Q and U. The polarization along the inner edges of the pixel is no 
longer visible at $p/\lambda<$1 as the correlation scale has increased from the multi-mode cases to
the single-mode case.

\begin{figure}[h]
\begin{center}
\includegraphics[width=6.5in]{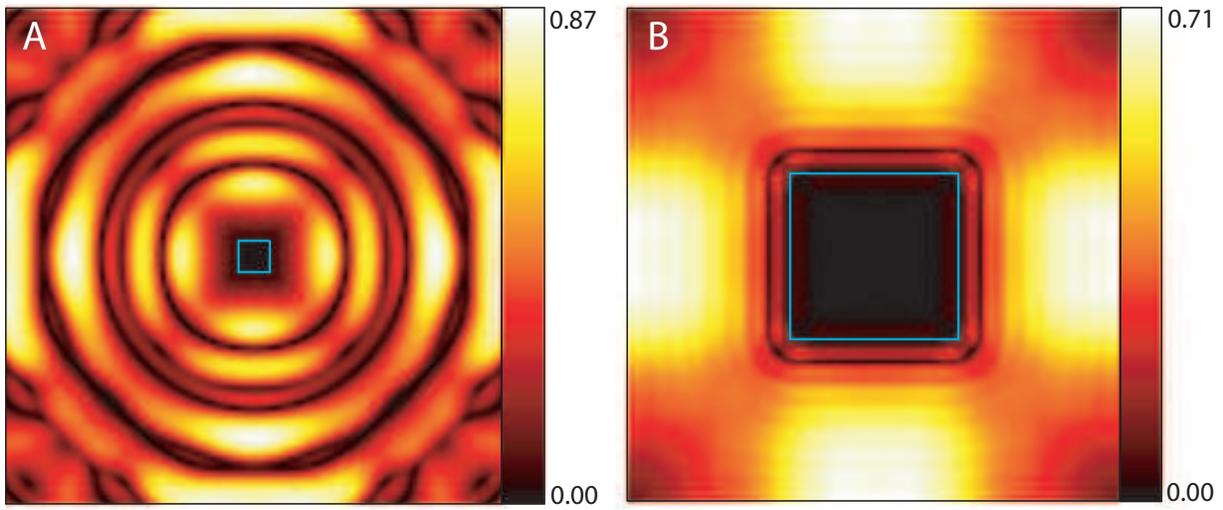}
\caption{The fractional polarization ($\sqrt{Q^2+U^2}/I$) as a function of position is shown for the cases
of $p/\lambda=0.25$ (A) and $p/\lambda=4.00$ (B). The blue square in each image shows the outline of the pixel
boundary used in each calculation.}
\label{fig:polmaps}
\end{center}
\end{figure}

Figure~\ref{fig:polmaps} shows the spatial distribution of fractional polarization $(\sqrt{Q^2+U^2}/I)$ for the extreme 
cases of $p/\lambda=0.25$ and $p/\lambda=4.00$. In both cases, the on-axis polarization is quite low, 
and the polarization of the scattered radiation quite high.  This is expected as scattering and
polarization are intimately related since both result from the same induced correlations.

\section{Pixel Cross-Coupling}

Perhaps the simplest and most important quantity for assessing the effect of 
$p/\lambda$ on a real detector array is the cross-coupling between adjacent pixels. By cross-coupling, 
we mean the the degree to which the emission from one small region of an incoherent 
sky contributes to the outputs of two detectors simultaneously. This is different from the 
problem of determining the correlations between the fluctuations in the outputs of two 
detectors, which is also possible using the model presented, but it is not what we have 
done here. The loss of resolution introduced by a telescope will serve to increase the 
overlap between the beam patterns of pixels, and thus in some sense, this model serves as a 
best case scenario for given a $p/\lambda$. In practice, for diffraction-limited systems, the 
imaging system should be designed such that the overlap due to the finite resolution of the 
telescope dominates. 

\begin{figure}[h]
\begin{center}
\includegraphics[scale=1]{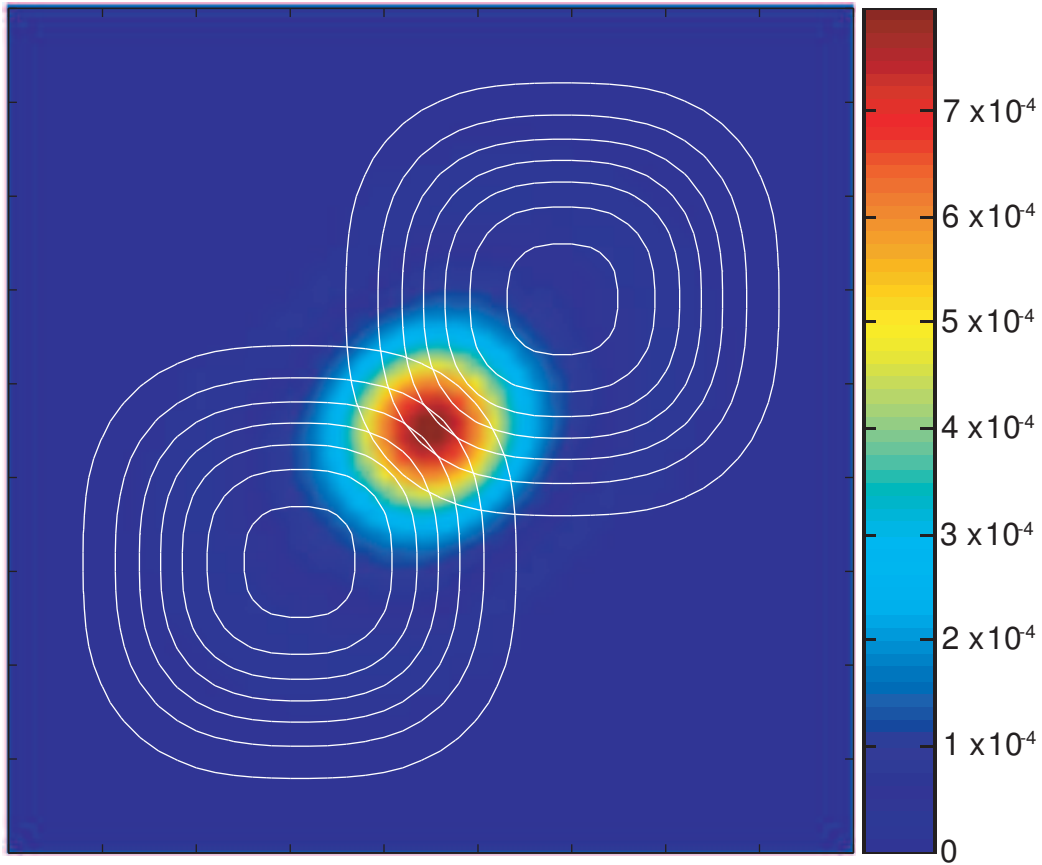}
\caption{
An example plot of the spatial correlation density between two pixels is shown. Here, one pixel is displaced
relative to the other by 1 pixel unit in each of the x and y directions ($\rho(x,y,1,1)$), and $p/\lambda=1$. Contours represent
the Stokes I power distributions for the two detectors and are set at the following fractions of the peak flux for each
pixel's beam: 0.14, 0.28, 0.42, 0.56, 0.70, 0.84, and 0.98. 
}
\label{fig:spatialcorr}
\end{center}
\end{figure}

\begin{figure}[h]
\begin{center}
\includegraphics[width=6.0in]{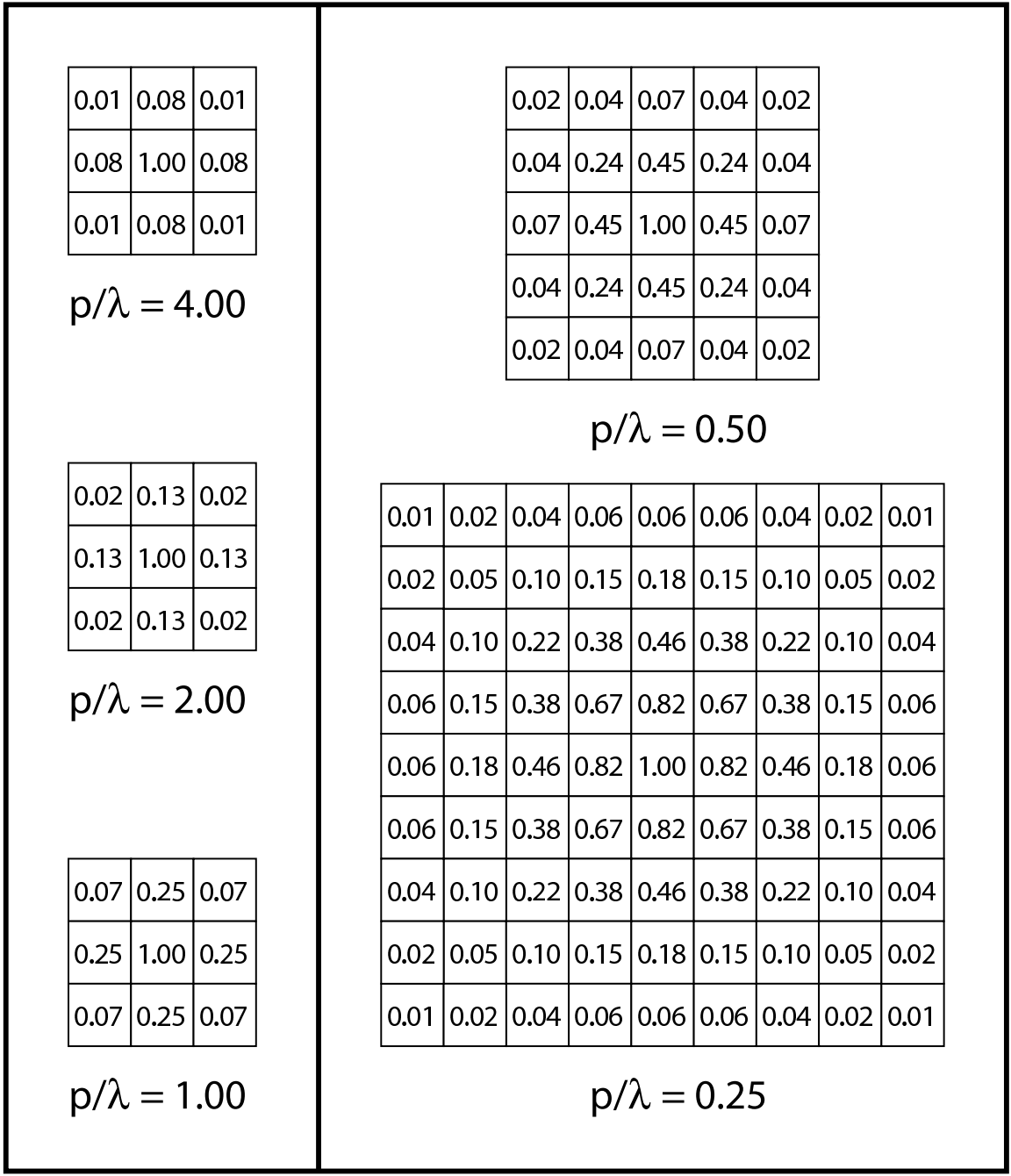}
\caption{For the five test cases of p/$\lambda$, we show the calculated overlap integrals ($\beta$) for pixels
with the illustrated spatial relationships to the central pixel (for which the overlap integral is
defined to be 1).
}
\label{fig:corr}
\end{center}
\end{figure}

We will define a cross-coupling factor, which is a function of position on the sky $(x,y)$, as 
well as the two-dimensional separation $(\Delta x,\Delta y)$ between the center of the 
two pixels being compared, as 
\begin{equation}
\rho(x,y,\Delta x, \Delta y)=\frac{1}{\Gamma}I(x,y)I(x+\Delta x, y+\Delta y)
\end{equation}
Here, $\Gamma=\int{[I(x^\prime,y^\prime)]^2dx^\prime\, dy^\prime}$, where the integral extends 
over all space. $\rho(x,y,\Delta x, \Delta y)$ is therefore a normalized cross-coupling factor.
Figure~\ref{fig:spatialcorr} shows an example of the cross-coupling factor for two pixels that 
are diagonal nearest neighbors. The total cross coupling factor for a uniform sky,
between two pixels whose beam patterns are separated by $\Delta x$ and $\Delta y$ in the 
$x$ and $y$ directions respectively, is simply
\begin{equation}
\beta(\Delta x, \Delta y)=\int{\rho(x,y,\Delta x, \Delta y)dx\,dy}
\end{equation}
The integral nominally extends over all space, but in reality is limited by the extent of our 
calculated beam profile models. We include a summary of the values of $\beta(\Delta x, \Delta y)$ 
for the cases studied previously in Fig.~\ref{fig:corr}.

\section{Analysis of Polarized Pixels}

With the interest in astronomical polarimetry growing, a logical extension of planar arrays is to pattern the
absorbers such that they are sensitive to linear polarization. Doing so allows the theoretical possibility of
stacking two orthogonally polarized pixels on top of one another in order to detect both modes of polarization 
simultaneously. 

\begin{figure}[h]
\begin{center}
\includegraphics[angle=-90,width=6.5in]{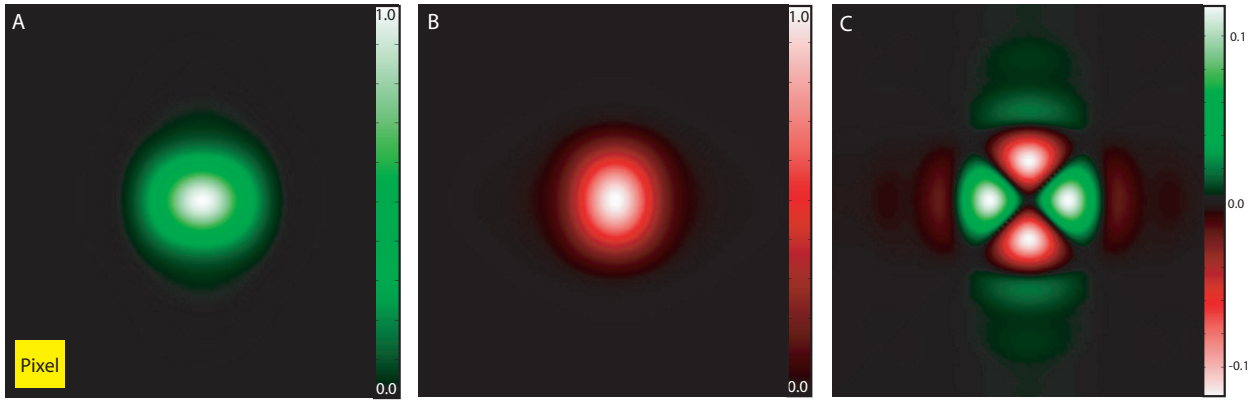}
\caption{The beam patterns for horizontally and vertically polarized pixels are shown in A and B, respectively.
The difference between the two is shown in C, which is analagous to the cross-polarization of the pixel. These calculations were done for $p/\lambda=0.5$. The size of a pixel is shown in the lower left corner
of panel A.
}
\label{fig:poldet}
\end{center}
\end{figure}

With minor variation (i.e. by eliminating 
all of the modes of one polarization from the blackbody source), the analysis technique described here can
be used to study the systematic effects one would expect when working in the small pixel limit.
Figure~\ref{fig:poldet} shows the intensity beam patterns for horizontally and vertically polarized detectors 
(A and B, respectively) for the case of $p/\lambda=0.5$. The difference between these two images, the 
incoherent analog of cross-polarization, is shown in (C). It is of interest to note that the effective beam 
size is smeared in the direction of polarization. This is consistent with the sign of Stokes Q in 
Figure~\ref{fig:Imaps}. One might expect that the polarization parallel to the pixel edge would be scattered 
more efficiently; however close to the pixel this is not the case. Looking further out in Figure~\ref{fig:poldet}
(A and B), one notices that the parallel polarization has more support in the highly scattered modes far from 
the pixel.

\section{Conclusion}

We have explored the consequences of using filled arrays of planar detectors in the limit where $p/\lambda$ is 
small, and found that in this limit, diffraction due to the edge currents in the pixels must be considered 
when designing optical systems. This diffraction has the effect of limiting the angular resolution of the 
instrument for a given plate scale. Thus, by carefully choosing the plate scale, it is possible to mitigate 
this effect. This tends to drive the design such that the pixels oversample the Airy function of the telescope. 
For polarized detectors in this same limit, systematic effects can become non-negligible, leading to cross-polarization 
that is in excess of 10\%.  

This model is idealized, but the correspondence of this model to a particular detector implementation could be improved. 
One such improvement involves focusing on the 
details of how the absorbing properties of the pixels are modeled. In our current implementation, the pixels 
are modeled as completely incoherent emitting/absorbing apertures. It is possible to tailor this formalism
to a specific coupling architecture by modeling the detailed material properties of the absorber. In addition, for a real array, the pixels
near the edge of the array will have different electromagnetic boundary conditions than those near the center.
This effect could also be included in the model.
 
The strength of this technique used in this work is in its ability to handle partially-coherent radiation
\citep{Withington03, Carter76}. Though
coherent analysis of
electromagnetic systems are useful and informative (eg. \cite{Wollack06} and references therein), the second-order statistical correlations introduced by these
systems are not directly accessible.  However, consideration of these correlations is essential
for understanding and optimizing the performance of submillimeter and millimeter astronomical systems.

\section*{Acknowledgments}
We would like to thank Lyman Page and Suzanne Staggs for their support of this work.

\clearpage

\begin{deluxetable}{l l l l l l l}
\tablecaption{A selection of both current and planned astronomical instruments that
employ or will employ arrays of planar bolometers\label{tab:instruments}}
\tablewidth{0pt}
\tablehead{
\colhead{Instrument} &
\colhead{Array Size} &
\colhead{Detector Type} &
\colhead{$\lambda$(mm)} &
\colhead{$p$(mm)} &
\colhead{$p/\lambda$}
}
\startdata
HAWC/SOFIA      &   12$\times$32 &  Semiconducting Bolometer & 0.053   &  1.00 &       18.9\\
                &   12$\times$32 &  Semiconducting Bolometer & 0.088   &  1.00 &       11.4\\
                &   12$\times$32 &  Semiconducting Bolometer & 0.155   &  1.00 &       6.5\\
                &   12$\times$32 &  Semiconducting Bolometer & 0.215   &  1.00 &       4.7\\\hline
SHARC II/CSO        &   12$\times$32 &  Semiconducting Bolometer & 0.350    &  1.00 &       2.9 \\
                &   12$\times$32 &  Semiconducting Bolometer &0.450     &  1.00 &       2.2 \\
                &   12$\times$32 &  Semiconducting Bolometer &  0.850   &  1.00 &       1.2 \\\hline
SCUBA 2/JCMT         &   64$\times$64 &  TES &       0.450   &  1.135&       2.5 \\
                &   32$\times$32 &  TES &       0.850   &  1.135&       1.3 \\\hline
PAR/GBT             &   8$\times$8 &  TES &  3.00    &  3.00 &       1.0 \\\hline
GISMO/IRAM           &   8$\times$16 &  TES &  2.00    &  2.00 &       1.0 \\\hline
ACT             &   32$\times$32 & TES &        1.13    &  1.00 &       0.9\\
                &   32$\times$32 & TES &        1.33    &  1.00 &       0.8\\
                &   32$\times$32 & TES &        2.07    &  1.00 &       0.5\\\hline
\enddata
\end{deluxetable}

\begin{deluxetable}{l c}
\tablecaption{A tabulation of the number of modes (along 1 dimension) used for each of the cases studied
\label{tab:usedmodes}}
\tablewidth{0pt}
\tablehead{
\colhead{p/$\lambda$} &
\colhead{Number of modes}
}
\startdata
0.25 &  8 \\
0.50 &  10 \\
1.00 &  12 \\
 2.00 &  14\\
4.00 &  16 \\
\enddata
\end{deluxetable}


\begin{thebibliography}{13}
\expandafter\ifx\csname natexlab\endcsname\relax\def\natexlab#1{#1}\fi

\bibitem[{Carter \& Somers(1976)}]{Carter76}
Carter, W.~H., \& Somers, L.~E. 1976, IEEE Trans. on Ant. and Prop., 24

\bibitem[{{Dicker} {et~al.}(2006){Dicker}, {Abrahams}, {Ade}, {Ames},
  {Benford}, {Chen}, {Chervenak}, {Devlin}, {Irwin}, {Korngut}, {Maher},
  {Mason}, {Mello}, {Moseley}, {Norrod}, {Shafer}, {Staguhn}, {Talley},
  {Tucker}, {Werner}, \& {White}}]{Dicker06}
{Dicker}, S.~R., {Abrahams}, J.~A., {Ade}, P.~A.~R., {Ames}, T.~J., {Benford},
  D.~J., {Chen}, T.~C., {Chervenak}, J.~A., {Devlin}, M.~J., {Irwin}, K.~D.,
  {Korngut}, P.~M., {Maher}, S., {Mason}, B.~S., {Mello}, M., {Moseley}, S.~H.,
  {Norrod}, R.~D., {Shafer}, R.~A., {Staguhn}, J.~G., {Talley}, D.~J.,
  {Tucker}, C., {Werner}, B.~A., \& {White}, S.~D. 2006, in Millimeter and
  Submillimeter Detectors and Instrumentation for Astronomy III. Edited by
  Jonas Zmuidzinas, Wayne S. Holland, Stafford Withington, and William D.
  Duncan. Proceedings of the SPIE, Volume 6275, pp. 627518 (2006).

\bibitem[{{Dowell} {et~al.}(2003){Dowell}, {Allen}, {Babu}, {Freund},
  {Gardner}, {Groseth}, {Jhabvala}, {Kovacs}, {Lis}, {Moseley}, {Phillips},
  {Silverberg}, {Voellmer}, \& {Yoshida}}]{Dowell03}
{Dowell}, C.~D., {Allen}, C.~A., {Babu}, R.~S., {Freund}, M.~M., {Gardner}, M.,
  {Groseth}, J., {Jhabvala}, M.~D., {Kovacs}, A., {Lis}, D.~C., {Moseley}, Jr.,
  S.~H., {Phillips}, T.~G., {Silverberg}, R.~F., {Voellmer}, G.~M., \&
  {Yoshida}, H. 2003, in Millimeter and Submillimeter Detectors for Astronomy.
  Edited by Phillips, Thomas G.; Zmuidzinas, Jonas. Proceedings of the SPIE,
  Volume 4855, pp. 73-87 (2003)., ed. T.~G. {Phillips} \& J.~{Zmuidzinas},
  73--87

\bibitem[{{Fowler}(2004)}]{Fowler04}
{Fowler}, J.~W. 2004, in Astronomical Structures and Mechanisms Technology.
  Edited by Antebi, Joseph; Lemke, Dietrich. Proceedings of the SPIE, Volume
  5498, pp. 1-10 (2004)., ed. J.~{Zmuidzinas}, W.~S. {Holland}, \&
  S.~{Withington}, 1--10

\bibitem[{{Harper} {et~al.}(2004){Harper}, {Bartels}, {Casey}, {Chuss},
  {Dotson}, {Evans}, {Heimsath}, {Hirsch}, {Knudsen}, {Loewenstein}, {Moseley},
  {Newcomb}, {Pernic}, {Rennick}, {Sandberg}, {Sandford}, {Savage},
  {Silverberg}, {Spotz}, {Voellmer}, {Waltz}, {Wang}, \& {Wirth}}]{Harper04}
{Harper}, D.~A., {Bartels}, A.~E., {Casey}, S.~C., {Chuss}, D.~T., {Dotson},
  J.~L., {Evans}, R., {Heimsath}, S., {Hirsch}, R.~A., {Knudsen}, S.,
  {Loewenstein}, R.~F., {Moseley}, S.~H., {Newcomb}, M., {Pernic}, R.~J.,
  {Rennick}, T.~S., {Sandberg}, E., {Sandford}, D.~B., {Savage}, M.~L.,
  {Silverberg}, R.~F., {Spotz}, R., {Voellmer}, G.~M., {Waltz}, P.~W., {Wang},
  S., \& {Wirth}, C. 2004, in UV and Gamma-Ray Space Telescope Systems. Edited
  by Hasinger, G{\"u}nther; Turner, Martin J. L. Proceedings of the SPIE,
  Volume 5492, pp. 1064-1073 (2004)., 1064--1073

\bibitem[{{Holland} {et~al.}(2006){Holland}, {MacIntosh}, {Fairley}, {Kelly},
  {Montgomery}, {Gostick}, {Atad-Ettedgui}, {Ellis}, {Robson}, {Hollister},
  {Woodcraft}, {Ade}, {Walker}, {Irwin}, {Hilton}, {Duncan}, {Reintsema},
  {Walton}, {Parkes}, {Dunare}, {Fich}, {Kycia}, {Halpern}, {Scott}, {Gibb},
  {Molnar}, {Chapin}, {Bintley}, {Craig}, {Chylek}, {Jenness}, {Economou}, \&
  {Davis}}]{Holland06}
{Holland}, W., {MacIntosh}, M., {Fairley}, A., {Kelly}, D., {Montgomery}, D.,
  {Gostick}, D., {Atad-Ettedgui}, E., {Ellis}, M., {Robson}, I., {Hollister},
  M., {Woodcraft}, A., {Ade}, P., {Walker}, I., {Irwin}, K., {Hilton}, G.,
  {Duncan}, W., {Reintsema}, C., {Walton}, A., {Parkes}, W., {Dunare}, C.,
  {Fich}, M., {Kycia}, J., {Halpern}, M., {Scott}, D., {Gibb}, A., {Molnar},
  J., {Chapin}, E., {Bintley}, D., {Craig}, S., {Chylek}, T., {Jenness}, T.,
  {Economou}, F., \& {Davis}, G. 2006, in Millimeter and Submillimeter
  Detectors and Instrumentation for Astronomy III. Edited by Jonas Zmuidzinas,
  Wayne S. Holland, Stafford Withington, and William D. Duncan. Proceedings of
  the SPIE, Volume 6275, pp. 627518 (2006).

\bibitem[{Holloway(1986)}]{Holloway86}
Holloway, H. 1986, Journal of Applied Physics, 60, 1091

\bibitem[{Keller(1962)}]{Keller62}
Keller, J. 1962, J. Opt. Soc. Am., 52, 116

\bibitem[{Mehta \& Wolf(1964)}]{Mehta64a}
Mehta, C.~L., \& Wolf, E. 1964, Phys. Rev., 134, A1143

\bibitem[{{Staguhn} {et~al.}(2006){Staguhn}, {Benford}, {Allen}, {Moseley},
  {Sharp}, {Ames}, {Brunswig}, {Chuss}, {Dwek}, {Maher}, {Marx}, {Miller},
  {Navarro}, \& {Wollack}}]{Staguhn06}
{Staguhn}, J.~G., {Benford}, D.~J., {Allen}, C.~A., {Moseley}, S.~H., {Sharp},
  E.~H., {Ames}, T.~J., {Brunswig}, W., {Chuss}, D.~T., {Dwek}, E., {Maher},
  S.~F., {Marx}, C.~T., {Miller}, T.~M., {Navarro}, S., \& {Wollack}, E.~J.
  2006, in Millimeter and Submillimeter Detectors and Instrumentation for
  Astronomy III. Edited by Jonas Zmuidzinas, Wayne S. Holland, Stafford
  Withington, and William D. Duncan. Proceedings of the SPIE, Volume 6275, pp.
  627518 (2006).

\bibitem[{Withington(2006)}]{Withington06}
Withington, S. 2006, in preparation

\bibitem[{Withington {et~al.}(2003)Withington, Tham, \& Yassin}]{Withington03}
Withington, S., Tham, C., \& Yassin, G. 2003, Proc. SPIE, 4855, 49

\bibitem[{Wollack {et~al.}(2006)Wollack, Chuss, \& Moseley}]{Wollack06}
Wollack, E., Chuss, D., \& Moseley, S. 2006, in Proc. of SPIE, ed.
  J.~Zmuidzinas, W.~Holland, S.~Withington, \& W.~Duncan, Vol. 6275

\end{thebibliography}
\end{document}